\documentclass[journal]{vgtc}                % final (journal style)
\ifpdf%                                % if we use pdflatex
  \pdfoutput=1\relax                   % create PDFs from pdfLaTeX
  \usepackage{graphicx}                % allow us to embed graphics files
  \DeclareGraphicsExtensions{.pdf,.png,.jpg,.jpeg} % for pdflatex we expect .pdf, .png, or .jpg files
\else%                                 % else we use pure latex
  \usepackage{graphicx}                % allow us to embed graphics files
  \DeclareGraphicsExtensions{.eps}     % for pure latex we expect eps files
\fi%

\graphicspath{{figures/}{pictures/}{images/}{./}} % where to search for the images

\usepackage{microtype}                 % use micro-typography (slightly more compact, better to read)
\PassOptionsToPackage{warn}{textcomp}  % to address font issues with \textrightarrow
\usepackage{textcomp}                  % use better special symbols
\usepackage{mathptmx}                  % use matching math font
\usepackage{times}                     % we use Times as the main font
\usepackage{cite}                      % needed to automatically sort the references
\usepackage{tabu}                      % only used for the table example
\usepackage{booktabs}                  % only used for the table example

\usepackage{balance}

\balance

\newcommand{\revision}[1]{\textcolor{black}{#1}}

\onlineid{0}

\vgtccategory{Research}
\vgtcpapertype{please specify}

\title{HeartBees: Visualizing Crowd Affects}

\author{Chao Ying Qin, Marios Constantinides, Luca Maria Aiello, and Daniele Quercia}
\authorfooter{
\item
 Chao Ying Qin is with Imperial College London, UK. E-mail: chaoying.qin@network.rca.ac.uk.
\item
 Marios Constantinides is with Nokia Bell Labs, Cambridge, UK. E-mail: marios.constantinides@nokia-bell-labs.com.
\item
 Luca Maria Aiello is with Nokia Bell Labs, Cambridge, UK. E-mail: luca.aiello@nokia-bell-labs.com.
 \item
 Daniele Quercia is with Nokia Bell Labs, Cambridge, UK. E-mail: daniele.quercia@nokia-bell-labs.com.
}

\shortauthortitle{Qin \MakeLowercase{\textit{et al.}}: HeartBees: Visualizing Crowd Affects}
\abstract{
Affective sharing within groups strengthens coordination and empathy, leads to better health outcomes, and increases productivity and performance. Existing tools for affective sharing face one main challenge: \emph{creating a representation of collective emotional states that is relatable and universally accessible}. To overcome this challenge, we propose HeartBees, a bio-feedback system for visualizing collective emotional states, which maps a multi-dimensional emotion model into a metaphorical visualization of flocks of birds. Grounded on Affective Computing literature and physiological sensing, we mapped physiological indicators that could be obtained from wearable devices into a multi-dimensional emotion model, which, in turn, our HeartBees can make use of. We evaluated our nature-inspired interactive system with 353 online participants, whose responses showed good consensus in the way they subjectively perceived the visualizations. Last, we discuss practical applications of HeartBees.
} % end of abstract


\CCScatlist{ % not used in journal version
 \CCScat{K.6.1}{Management of Computing and Information Systems}%
{Project and People Management}{Life Cycle};
 \CCScat{K.7.m}{The Computing Profession}{Miscellaneous}{Ethics}
}

\teaser{
  \centering
  \includegraphics[width=0.9\linewidth]{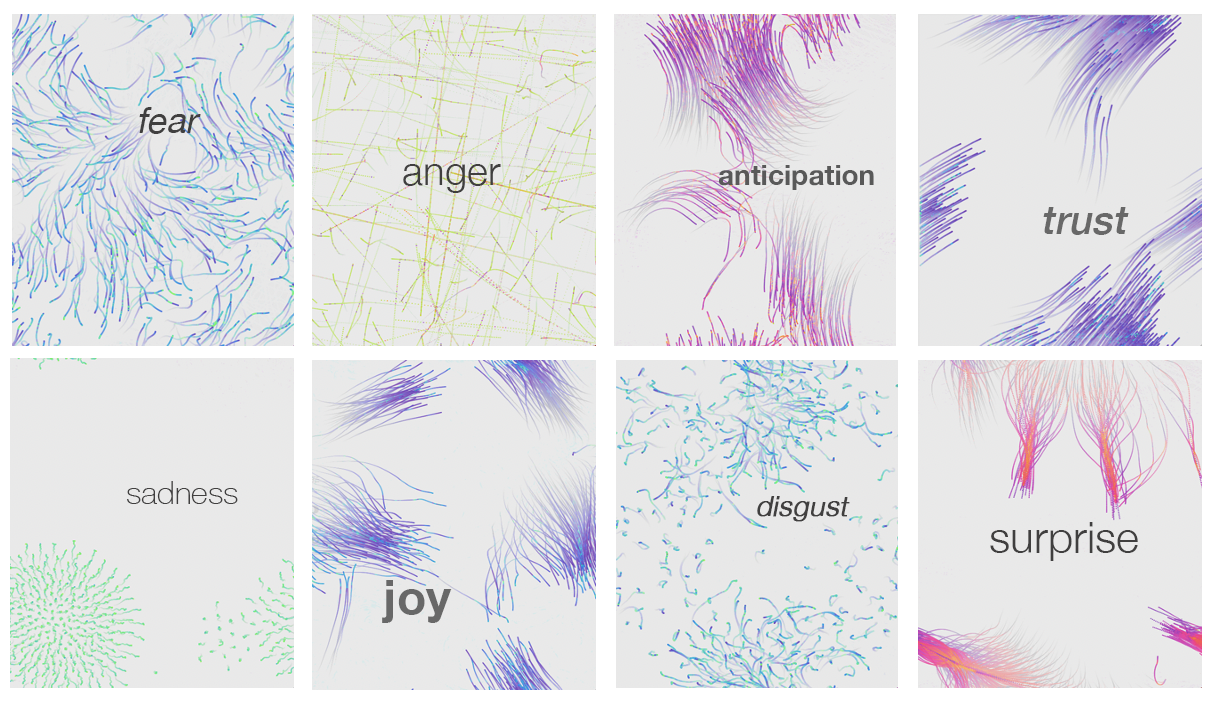}
  \caption{HeartBees. A metaphorical visualization  that uses the movements of a flock of birds in a plane to represent eight basic emotions: fear, anger, anticipation, trust, surprise, disgust, joy, and sadness.}
  \label{fig:teaser}
}

\vgtcinsertpkg

\usepackage{graphicx}
\usepackage{subcaption}
\usepackage{multirow}
\usepackage{makecell}

\begin{document}



\maketitle

%% \section{Introduction} %for journal use above \firstsection{..} instead

\section{Introduction}
\label{sec:section1}
Emotional Intelligence (EI) is the ability of individuals to understand and recognize their own and other people's emotions in such a way that will help them alleviate stress and overcome challenges, communicate effectively, neutralize conflicts, and empathize with others~\cite{salovey1990emotional}. At individual level, developing EI helps building strong and lasting relationships, succeed at work, and, ultimately, improve at every aspect of their life. Emotional Intelligence is also an emergent property of groups. Research in the area of organizational behavior has accumulated abundant evidence that affective sharing within groups activates a positive self-reinforcing spiral that strengthens coordination and empathy among members~\cite{walter2008positive}. Supporting emotional intelligence at the collective level is therefore important to increase both people's well-being and their chances of success when cooperating. This objective poses one main challenge: one needs to create a representation of the collective emotional state that is relatable and universally accessible.
% \footnote{http://www.thelisapark.com/blooming}
To overcome this challenge, since the 70s, bio-feedback has been a widely employed mind-body technique to help people gain control over involuntary bodily functions~\cite{hauri1975biofeedback}. Traditionally, bio-feedback systems turn bodily signals (e.g., heart rate) into readable representations that can increase awareness of the connection between physiological states and the body's inner functioning. These systems often use numerical or simple graphical representations such as heartbeats (bpm) or electrocardiography waveforms. While these representations are easily comprehensible among the medical practitioners, they are not always easily interpretable by the general audience~\cite{costa2016emotioncheck}. An alternative way of conveying people's physiological signals is the use of metaphorical visualizations. This type of visualization creates an analogy between characteristics of well-understood, and often universally acceptable images or patterns, and a more poorly understood or complex data source~\cite{ziemkiewicz2008shaping}; our emotional states is one of them. It is also well-known that the use of metaphors and analogies are key aspects of human cognition~\cite{boicho2001analogical}, which enable humans to understand abstract information with familiar objects such as images or simulations~\cite{blackwell2006reification}. For example, previous systems made use of tangible artifacts that mimic real-world objects~\cite{yu2017stresstree}, light~\cite{yu2018delight}, or movement. Our contribution draws inspiration from this recent line of research. More importantly, the HCI and Ubiquitous Computing research has worked on tools that will help gather knowledge of psychological states from bodily signals at users' everyday environment, beyond medical facilities. In turn, these signals could be translated into our emotional states that bio-feedback systems can make use of. Recent advances in wearable sensing could facilitate systematic monitoring of large populations~\cite{hansel2016large,aiello2018hearts,schmidt2019multi} to estimate collective emotional states through machine-learning algorithms~\cite{park2020wellbeat, gjoreski2017monitoring,gloor2018aristotle,quiroz2018emotion}. By applying such models, for example, the Inbodied Interaction paradigm~\cite{schraefel2019in5} foresees the design of interactive systems that listen and adapt to people's bodily signals.

Specifically, we aim to develop a visualization that maps a multi-dimensional emotion model, and uses a nature-inspired metaphor to visualize collective emotions of crowds. In so doing, we made two main contributions:

\begin{itemize}
    \item We developed HeartBees (\S\ref{sec:section4}), a web-based bio-feedback system inspired by the Boids model. It maps a multi-dimensional emotion model in a metaphorical visualization of flocks of birds moving on a plane. In particular, it translates the eight primary emotions from Plutchik's theoretical model of emotions into motion configurations that control the movements of the Boids in the space. Grounded on Affective Computing and physiological sensing literature, we mapped physiological indicators that our HeartBees could make use of.
    \item We evaluated our system's ability to convey people's emotional states through crowdsourcing (\S\ref{sec:section5}). Our results showed that certain emotions, especially negative ones, were perceived as one would expect, and they were so by a variety of people.
\end{itemize}

\section{Related Work}
\label{sec:section2}
Our work relies on principles from Affective Computing, HCI, and InfoVis, which we describe next. 

\subsection{Affective Computing}
The ability of machines to recognize and interpret human affects is the subject of Affective Computing~\cite{picard2000affective}. This branch of computing quantifies people's emotional state into discrete categories based on patterns discovered in various signals and cues such as facial, auditory, textual, gestures or bodily, and physiological signals. In human emotion literature, three most widely used schemas that describe a handful of basic emotions derived from biological responses are the Ekman's~\cite{ekman1992argument} six emotion model, Plutchik's \cite{plutchik1984emotions} eight emotion categorization, and the positive-negative activation (PANAS) model of emotion created by Watson and Tellegen~\cite{watson1985toward}, which suggests that positive and negative affect are two separate systems. By employing such models, the scientific community proposed solutions that leverage bodily signals from wearable sensors and mobile phones~\cite{sano2013stress} to detect stressful situations, regulate our emotional state~\cite{costa2016emotioncheck}, analyzed auditory input~\cite{costa2018regulating}, textual sentiment in online discussions~\cite{choi2020ten, saldias2019tweet}, and facial cues from video streams~\cite{nicolaou2011continuous}.

% \footnote{https://www.affectiva.com/}
% \footnote{https://www.myfeel.co/}
% \footnote{https://ouraring.com/}
% \footnote{https://moodmetric.com/}
In practice, many consumer and enterprise products have been developed to offer behavioral suggestions for self-improvement. Affectiva employs facial video analysis to generate percentage representations of discrete emotional categories such as joy, anger, and fear to help businesses understand how their customers feel. Products such as Feel, Oura, and Moodmetric, to name a few, capture emotional states through physiological indicators (e.g., heart rate, skin conductance, EEG) to generate quantitative or categorical representations of emotional states, and recommend behavioral interventions to improve emotional wellness.

Among the various physiological indicators that are linked to our psychological and emotional states, heart rate variability (HRV) is a widely used one. HRV is the physiological phenomenon that describes the variation in the time interval between heartbeats, and has been found to be a promising physiological marker to assess humans' Autonomic Nervous System (ANS). In human body, the ANS is responsible for controlling body functions that are not consciously directed, such as the heartbeat or breathing~\cite{camm1996heart}. The ANS is divided into the sympathetic and parasympathetic nervous system. The former prepares the body for intense physical activity (`fight-or-flight'), while the latter prepares the body for relaxation (`rest and digest')~\cite{camm1996heart}. By capturing the activity of both components of the ANS, one can assess how well a person responds to physical or emotional stimuli~\cite{castaldo2015acute}. Today, wearable devices are fully equipped with `body-sensors' such as motion, heart rate, skin conductance, and temperature sensors make it possible to gather complex physical measurements outside medical facilities to understand people's emotional states in free-living conditions. Their widespread adoption allows systematic monitoring of large populations~\cite{hansel2016large,aiello2018hearts,schmidt2019multi} that can be used to estimate collective emotional states through machine-learning algorithms~\cite{park2020wellbeat, gjoreski2017monitoring,gloor2018aristotle,quiroz2018emotion}. Quiroz et al.~\cite{quiroz2018emotion} used smartwatch readings from accelerometers and gyroscopes to detect changes in the way an individual walks, which, in turn, reflects that individual's current mood. Gjoreski et al.~\cite{gjoreski2017monitoring} used Empatica devices to detect stress by combining HRV and electrodermal activity analysis in controlled laboratory conditions, and then applied that knowledge in real-life data. Gloor et al.~\cite{gloor2018aristotle} showed that happiness is strongly linked with intense activity by combining data obtained from smartwatches including acceleration, heart rate, light level, and location.

\begin{figure}
  \centering 
\begin{subfigure}{0.3\linewidth}
  \centering
  \includegraphics[width=\linewidth]{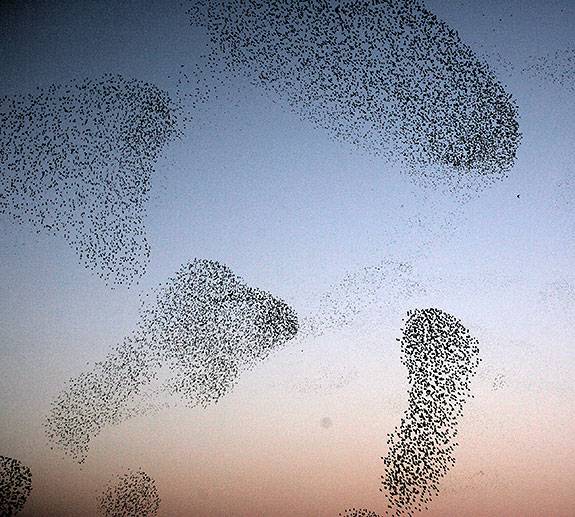}
\end{subfigure}
\begin{subfigure}{0.3\linewidth}
  \centering
  \includegraphics[width=\linewidth]{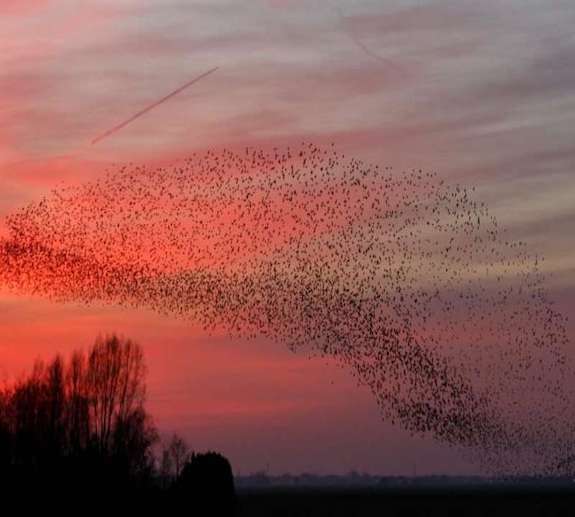}
\end{subfigure}
\begin{subfigure}{0.3\linewidth}
\centering
  \includegraphics[width=\linewidth]{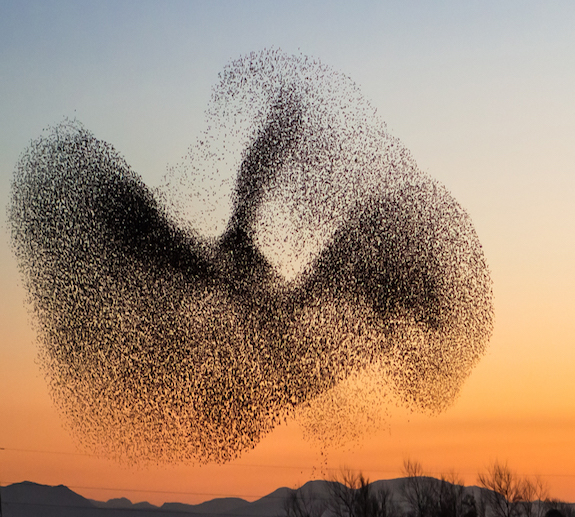}
\end{subfigure}
\caption{Starling murmurations: simple rules, complex patterns. From left to right as the birds swoop~\cite{starlings}, create unique yet complex patterns. These patterns might reflect the complexities of our collective emotional states.}
\label{fig:starlings}
\end{figure}

% \protect\footnotemark
% \footnotetext{https://phys.org/news/2019-02-starling-murmurations-science-nature-greatest.html}

\subsection{Design for Affective Sharing}
To communicate people's emotional state, researchers and practitioners have long before resorted to various bio-feeback systems that convey emotional states into readable representations such as numerical, graphical, material, or artistic, with the aim of increasing awareness of the connection between physiological states and the body's inner functioning~\cite{hauri1975biofeedback}. We draw inspiration from various lines of research including biophilic design, the use of metaphors and technological artifacts in the design, and material data representations.

\vspace{4pt}\noindent\textbf{Biophilic Design.} `Biophilia' is humans' inherent connection to nature; our love of life and the natural world~\cite{wilson}. Biophilic design creates spaces, objects, and technologies that provide a natural and pleasant environment for people to live and work. Contact with natural elements was found to increase the productivity of workers and to decrease mental fatigue
~\cite{nieuwenhuis2014relative}. StressTree~\cite{yu2017stresstree} is an example of bio-feedback system that uses a stylized tree to assist  people in relaxation training and stress management by changing shape and size based on people's heart rate variability.

% \footnote{http://www.thelisapark.com/blooming}
\vspace{4pt}\noindent\textbf{Metaphors and Technological artifacts.} Metaphorical visualizations aim to present data in a more evocative, meaningful, and thought provoking way to help an individual or a group of people grasp the underlying complexities of data. They create an analogy between characteristics of well-understood, and often universally acceptable images or patterns, and a more poorly understood or complex data source~\cite{ziemkiewicz2008shaping}; our emotional states is one of them. It is also well-known that the use of metaphors and analogies are key aspects of human cognition~\cite{boicho2001analogical}, which enable humans to understand abstract information with familiar objects such as images or simulations~\cite{blackwell2006reification}. Examples include the use of tangible artifacts that mimic real-world objects~\cite{yu2017stresstree}, light~\cite{yu2018delight}, or movement~\cite{lisapark}. Other systems resort to casual visualizations that become more playful (gamification) or more artistic. For example, Chill-Out is a bio-feedback game for relaxation training that maps a user's breathing rate into a game difficulty~\cite{parnandi2013chill}, while Cardiomorphologies offers an abstract visual artwork by mapping heartbeat and breath into a series of colorful rings~\cite{muller2006creating}. Inspiration could also be drawn by observing and studying animals behaviour. For example, murmuration is the phenomenon that results when hundreds, or even thousands, of starlings fly in swooping, creating coordinated patterns through the sky~\cite{young2013starling}. While these collective behaviors exhibit characteristics of self-organization, simple repeated interactions yield complex emergent patterns~\cite{sumpter2010collective, camazine2003self}; patterns that might reflect the complexity of our emotional states (Figure~\ref{fig:starlings}).

\vspace{4pt}\noindent\textbf{Material data representations.} This breed of representations make use of recent advances in digital fabrication, tangible interfaces, and shape-changing displays to offer data physicalizations that support embodied interpretation of our emotional states~\cite{jansen2015opportunities}. For example, Biolesce~\cite{biolesce} is a series of iterative installations and sculptures that display the viewer's heart rate in real-time through bioluminescent algae. Others include the use of fabric material as data displays. For example, Devendorf et al.~\cite{devendorf2016don} created Ebb, a slow, color-changing novel textile display that evokes personal style associations. In a similar vein, Howell et al.~\cite{howell2016biosignals} developed Hint, a thermochromic t-shirt that changes color based on wearer's skin conductance levels to understand how people react under different conversations.

% \footnote{http://www.tylersfox.com/487}
\section{Research Goal}
\label{sec:section3}
Our broad research goal is to build a bio-feedback system for visualizing crowd affects. Drawing inspiration from various lines of research in HCI and InfoVis, as stated in the previous section, we set out to explore the underlying mapping of a multi-dimensional model into a metaphor that conveys the complexity of our emotional states, and explore its effectiveness. \revision{Therefore, developing a visualization for sharing emotional states is a fundamental step towards building an end-to-end bio-feedback system.} Our research questions is:

\emph{RQ}: How to create a visual metaphor of basic emotions that is relatable and universally accessible?

We first motivate our design decisions and implementation of our system, and we then evaluate it in a crowdsourcing deployment study. 

\section{HeartBees}
\label{sec:section4}
We introduce HeartBees (http://social-dynamics.net/heartbees), a web-based tool for visualizing collective emotions based on the Boids model. We first motivate our choice of the Boids model, and we then present the basic mechanics of model and describe how we mapped emotional states to specific boids configurations. 

\subsection{Why Boids?}
Behavior models simulate the coordinated behavior in human society wherein each member relies on the decisions of others to guide his or her own actions.
The Boids model is a type of behavior model that offers a simple way to depict group dynamics that resonates with the human instinctive interpretation of collective behavior~\cite{schraefel2019in5}. Craig Reynolds described it as an approach to simulate the aggregate motion of a flock of birds, a herd of land animals, a school of fish, or a swarm of bees through a distributed behavioral model~\cite{reynolds1987flocks}. As we previously discussed (\S\ref{sec:section2}), simple repeated interactions in animals' behaviour can produce complex emergent patterns in which a metaphor could be drawn and mapped to our emotional states.

We used this model as an interactive bio-feedback system to simulate the aggregate physiological behavior of a group of people, in a playful way. Particles that move in a life-like fashion could stimulate what anthropologists and sociologists have called `collective effervescence'; the synchronicity in thoughts, actions, or emotions in groups and societies~\cite{xygalatas}. There is substantial scientific evidence that exposure to synchronicity increases the potential to create positive emotions that weaken interpersonal boundaries~\cite{wiltermuth, konvalinkaa}.

\begin{figure}
  \centering 
\begin{subfigure}{0.3\linewidth}
\centering
  \includegraphics[width=\linewidth]{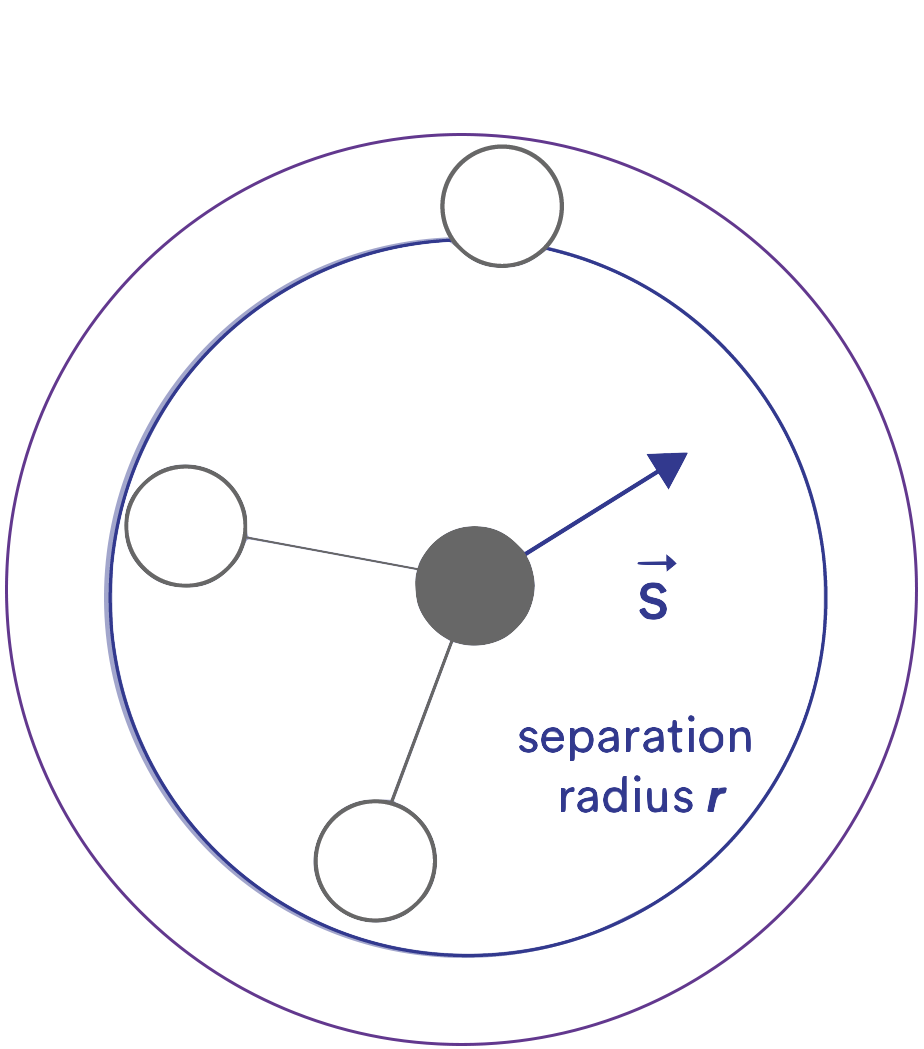}
  \caption{separation}
  \label{fig:separation}
\end{subfigure}
\begin{subfigure}{0.3\linewidth}
  \centering
  \includegraphics[width=\linewidth]{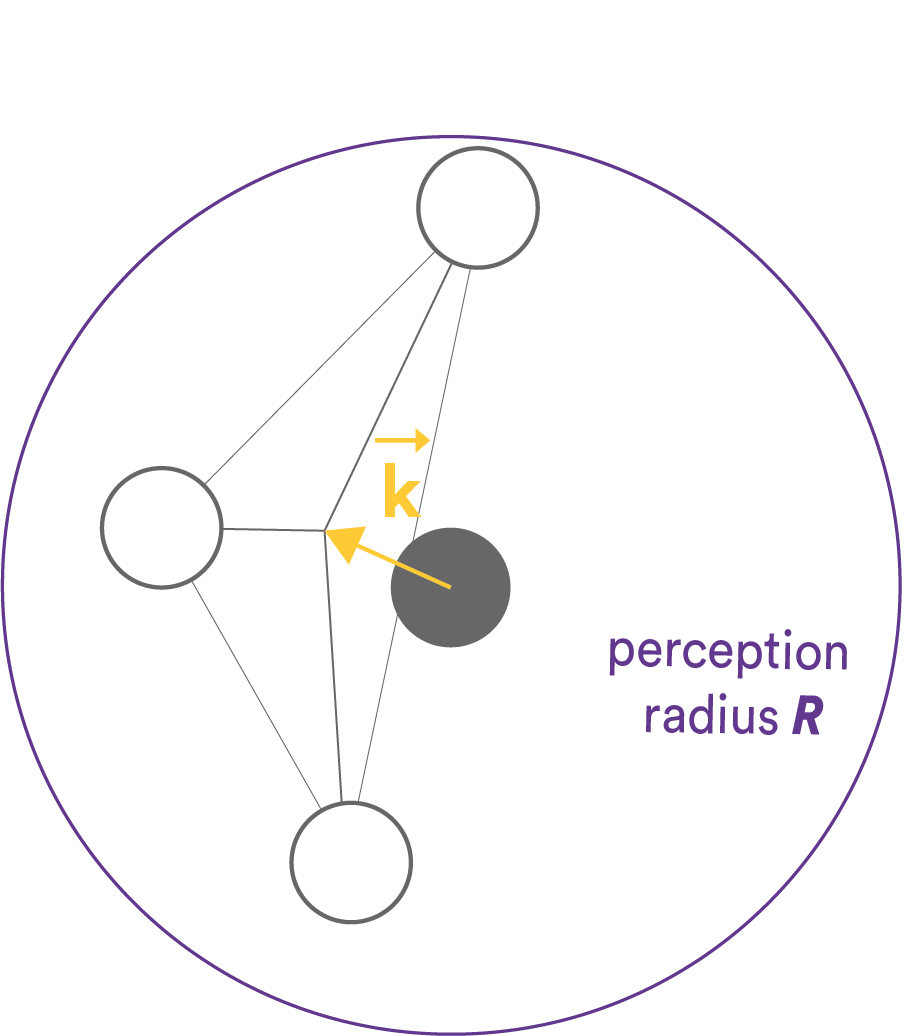}
  \caption{cohesion}
  \label{fig:cohesion}
\end{subfigure}
\begin{subfigure}{0.3\linewidth}
\centering
  \includegraphics[width=\linewidth]{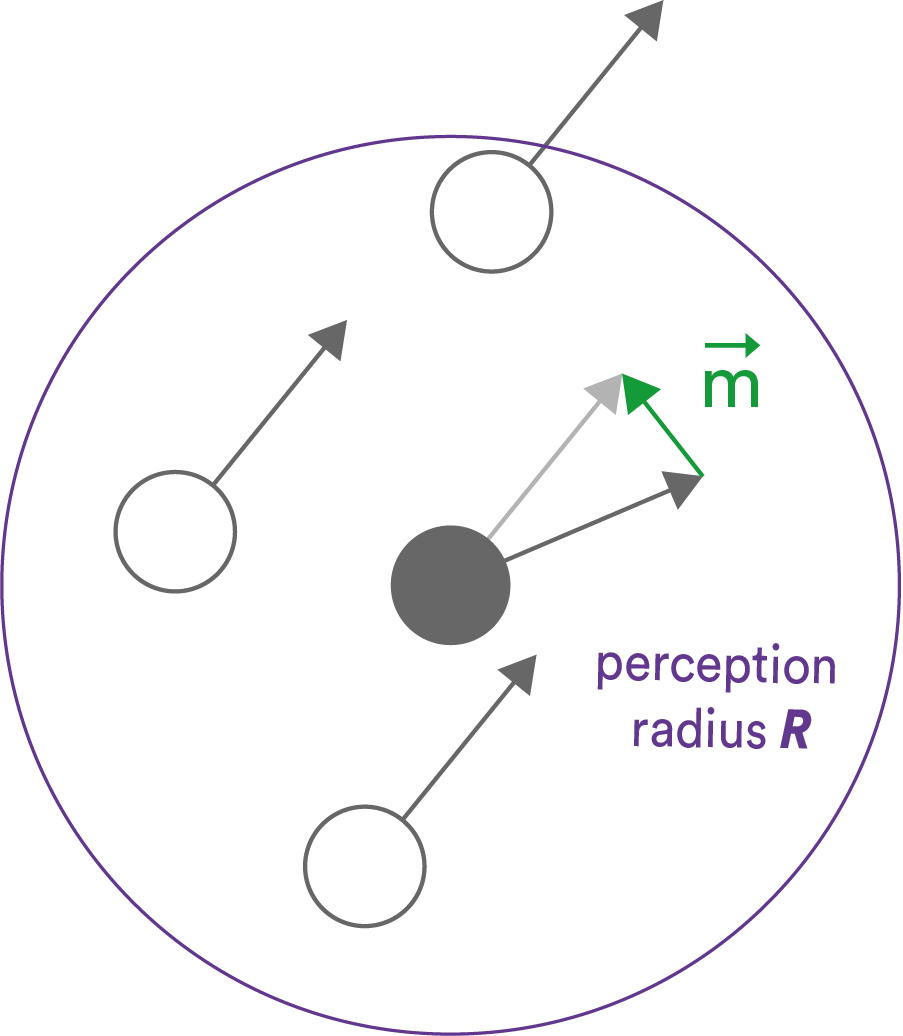}
  \caption{alignment}
  \label{fig:alignment}
\end{subfigure}

\caption{The underlying mechanics of Reynold's Boids model~\cite{reynolds1987flocks}. A flock of boids simulates a life-like collective motion based on a combination of three rules: (a) separation, (b) cohesion, and (c) alignment.}
\label{fig:boids_model}
\end{figure}

\begin{table}
\caption{Our mapping of Plutchik's multi-dimensional emotion model into boids configurations.}
\label{tbl:boids_coefficients}
\centering
\begin{tabular}{l|llllll}
    & $S$ & $M$ & $K$ & R & r & V\\ \hline\hline
joy          & 0.05      & 0.05     & 0.05       & 60 & 30 & 2 \\
sadness      & 0.05      & 0.05     & 0.05       &  0 & 30 & 1  \\\hline
fear         &   0.1        &   0.05       &   0.05  & 60 & 30 & 1   \\
anger        &   0.01       &  0.1         &   0.1       &      0 & 0   & 10 \\\hline
trust       & 0.05    &    0.1     &   0.05         & 60 & 30 & 2 \\
disgust      &   0.1        &    0.05    &  0.1          &   60  & 60& 5\\\hline
surprise    &    0.05    &    0.05     &   0.1          &  60   & 60 & 5\\
anticipation     &    0.1        &  0.1      &     0.05       &  60  & 30 & 4    \\\hline
\end{tabular}
\end{table}

\subsection{Model parameters}
\begin{figure*}
  \centering 
\includegraphics[width=0.89\linewidth]{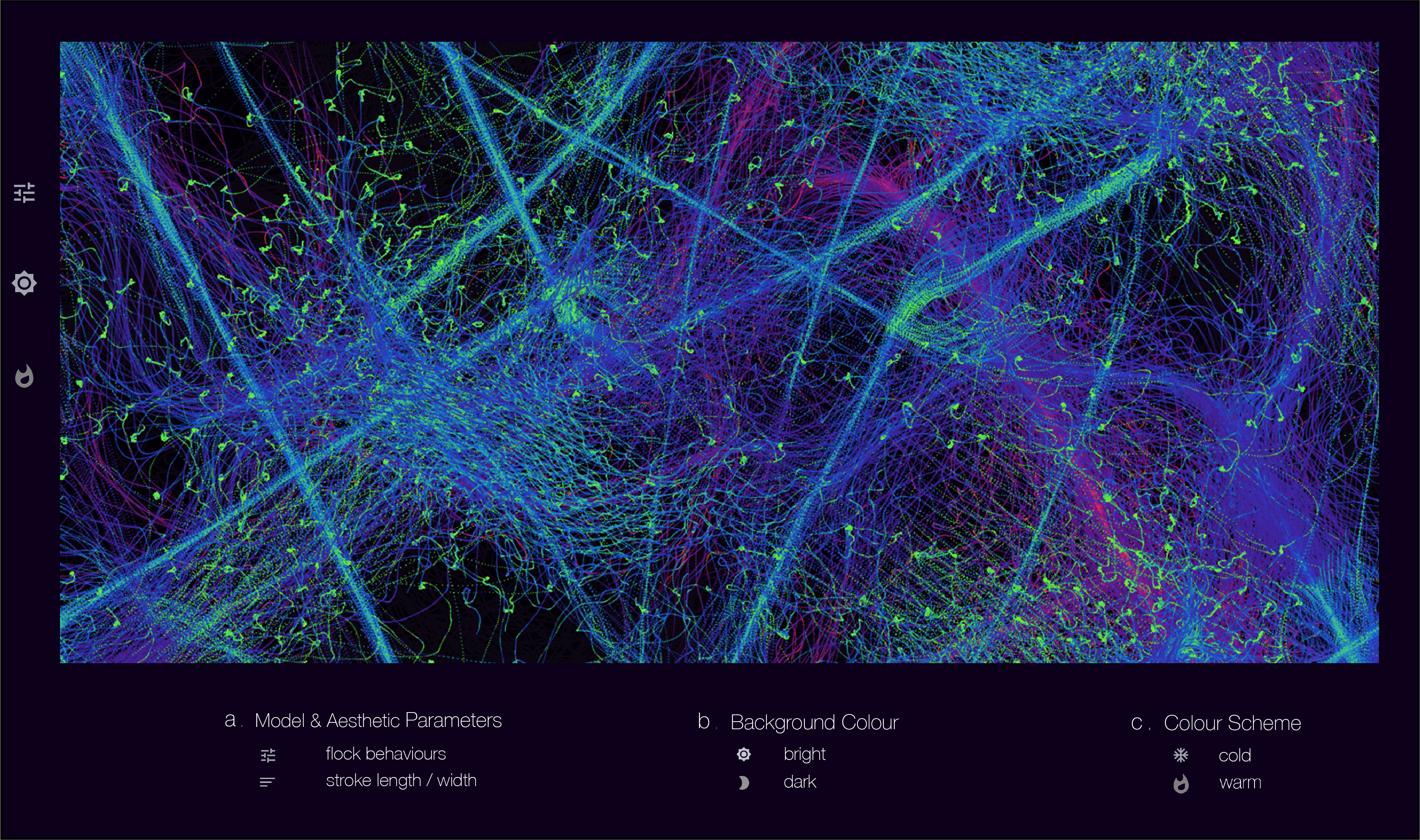}

\caption{HeartBees User Interface. Aesthetics parameters (i.e., stroke length and width), canvas background color (i.e., bright and dark), and boids color scheme (i.e., cold and warm) can be manually configured. The example HeartBees visual was generated by setting the stroke length to 100, while the stroke width set to 30. Both cold and warm color schemes were used on the boids drawn on a dark background color canvas.}
\label{fig:heartbees_interface}
\end{figure*}

The Boids model is a collection of independent particles (called `boids') that follow physical rules determined by their local perception of the environment and their neighbors; such behaviors include wandering, arriving, pursuing, fleeing, and evading~\cite{reynolds1999steering}. A flock of $N$ boids simulates a life-like collective motion based on a combination of three rules: separation $S$, alignment $M$, and cohesion $K$, as illustrated in Figure~\ref{fig:boids_model}, and can be configured in our interface (Figure~\ref{fig:heartbees_interface}a). Separation defines how boids avoid collisions with nearby flockmates, alignment attempts to match a boid's velocity with that of neighbors, and cohesion enables boids to stay closer to nearby flockmates. In addition to these dimensions, the model is initialized with three additional parameters: the particle velocity $V$; the range of perception $R$, which is used as a proximity threshold to trigger alignment and cohesion forces; and a range of separation r, which determines the minimum distance between boids. We implemented the Boids visualization as a web-based application using HTML5/JavaScript and D3~\cite{bostock_d3_2011}.

\subsection{Aesthetics}
In addition to the model parameters, our HeartBees web-based application allows users to control the aesthetic output of the visualization (Figure~\ref{fig:heartbees_interface}). \revision{Drawing inspiration from fractal patterns observed in nature~\cite{taylor2002order}, and color psychology~\cite{whitfield1990color}, our visualization encodes these patterns into colored Boids.} In particular, one can control the boids' \emph{stroke length} and \emph{stroke width} (Figure~\ref{fig:heartbees_interface}a) \revision{as a way to create an analogy between the Boids movements and the complex patterns observed in nature; previous work found that fractal patterns are aesthetically pleasing and help reduce stress levels~\cite{taylor2006reduction}}. The stroke length describes the trail of the boids as time passes; the higher its value, the longer the light ray prolongs. When the stroke length set to 100, all movement trails are preserved on the canvas. The stroke width describes the thickness of the colored paint strokes that traces the boids' movement. Furthermore, users can change the canvas background color in which the boids are being drawn (Figure~\ref{fig:heartbees_interface}b). \revision{Taking inspiration from color psychology~\cite{whitfield1990color}, we aimed at providing coloring options that are universally relatable.} Two options are available: (a) a dark background (dark-blue color), and (b) a bright background (white-grey color). Finally, the boids color scheme can be set to cold, warm, or a combination of the two (Figure~\ref{fig:heartbees_interface}c). The warm color palette includes combinations of orange, red, and yellow colors, while the cold color palette consists of green, blue, indigo and violet colors.

\subsection{Mapping emotions to Boids' motion configurations}
Inspired by the collective behavior of animals, we combined the Boids model parameters to create visual patterns that aim at conveying the eight basic emotions in Plutchik's model: Joy, Sadness, Fear, Anger, Trust, Disgust, Surprise, and Anticipation (Figure~\ref{fig:teaser}). We generated the mapping of flock behavior by varying the coefficients of the three forces, the maximum velocity, the separation radius, and the range of perception. The full list of coefficients is presented in Table~\ref{tbl:boids_coefficients}. For example, by increasing the degree of alignment the flock stays together, thus giving a sense of trust or joy, and conveying positive emotions. On the contrary, while keeping the degree of separation and cohesion constant, members of the flock separate from each other, thus giving a flavor of negative sensation, and conveying negative emotions such as disgust or fear. Additionally, we used the speed of motion to express ``energy'' exhibited by the flock. For example, anger and anticipation will be very fast motion, while sadness and fear are in the lowest energy spectrum.

\begin{figure*}
    \centering 
    \includegraphics[width=\linewidth]{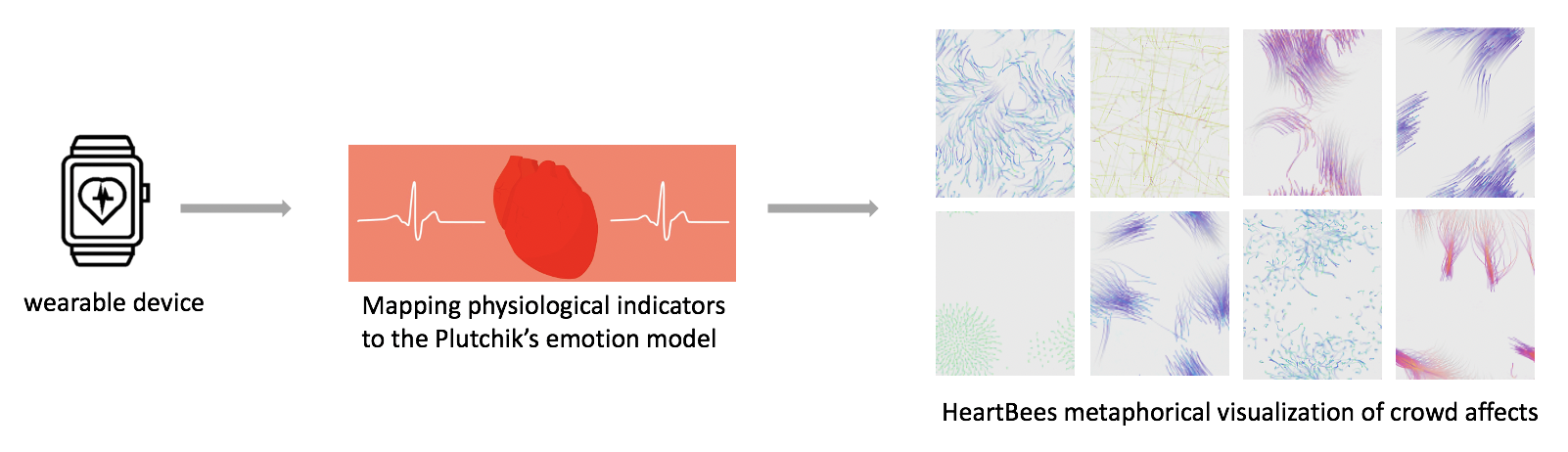}
    \caption{A schematic overview of HeartBees. Physiological data obtained from wearable devices can be mapped into a multi-dimensional emotion model, which in turn HeartBees translates into a metaphorical visualization of a flock of birds.}
    \label{fig:overview_diagram}
\end{figure*}

\begin{table}
\centering
  \caption{Mapping HR and HRV parameters (RMSSD and LF/HF) into the eight basic Plutchik's emotions. Adapted version from~\cite{kreibig2010autonomic}.}
  \label{tab:mapping_emotions}
  \begin{tabular}{cccc}
    \toprule
    
    & \textbf{HR} & \textbf{RMSSD} & \textbf{LF/HF}\\
    \midrule
    Joy & $\uparrow$ & $\downarrow$ & $\downarrow$ \\
    Disgust &  $\uparrow$ &  $\uparrow$ &  $\uparrow$\\
    Trust & $\uparrow$ &  $\uparrow$ & $\downarrow$\\
    Anger & $\uparrow$ & $\downarrow$ & $\uparrow$ \\
    Surprise & $\uparrow$ & $\uparrow$ & $\downarrow$\\
    Anticipation & $\downarrow$ & $\uparrow$ & $\downarrow$ \\
    Sadness & $\downarrow$ & $\downarrow$ & $\uparrow$\\
    Fear & $\downarrow$ & $\downarrow$ & $\uparrow$\\
    \bottomrule
  \end{tabular}
\end{table}

\subsection{Mapping physiological indicators to emotions}
Over the last decades, research in Affective Computing has accumulated evidence that links various types of signals to people's emotional states. As previously discussed (\S\ref{sec:section2}) these include facial expressions, voice intonation, textual sentiment, or bodily signal obtained from mobile and wearable sensors. While any of these datastreams could be possibly used to detect people's emotional states and, in turn, serve as an input to our HeartBees, we focused on HRV, a promising and widely used physiological indicator that is available thanks to recent advanced in wearable sensing. HRV provides an assessment of our Autonomic Nervous System; for example, one could obtain an assessment whether an individual is stressed or not, feeling happy or frustrated. Grounded on previous reviews on physiology and emotional states, we provide the reader a mapping of heart rate and heart rate variability parameters into Plutchik's multi-dimensional model, and visually illustrate how this type of physiological data can serve our HeartBees prototype (Figure~\ref{fig:overview_diagram}). 

Kreibig et al.~\cite{kreibig2010autonomic} provide an extensive review of these findings that link HRV parameters to our emotional states. Building on this review, we grounded our mapping of physiological indicators to the eight emotional states. For example, previous studies have shown that people feeling joy or other strong positive emotions usually experience an increase in HR and, at the same time, a decrease in both RMSSD and LF/HF. On the contrary, HR and RMSSD decrease and LF/HF increases when experiencing negative emotions such as sadness or fear. 

Based on the relevant previous literature, we found that HR, RMSSD, and LF/HF in combination can set apart different emotions quite effectively. By considering a coarse distinction of high vs. low range of the signals, we summarized the mapping between these three metrics and the eight emotions in Table~\ref{tab:mapping_emotions}. According to the literature, four dimensions (joy, disgust, anger, anticipation) are characterized by unique footprints of high/low levels of HR, RMSSD, and LF/HF. Two pairs of emotions, namely trust and surprise, and sadness and fear, have similar footprints; a not surprising observation given the complexities of our emotional states. 

\revision{To integrate the heart rate sensors (Figure~\ref{fig:overview_diagram}), we built on top of our previous work in which we demonstrated the feasibility of obtaining, processing, and analyzing physiological data from smartwatches~\cite{heartbees_mobilehci20}.}

\section{Evaluation}
\label{sec:section5}
To evaluate HeartBees, we investigated how people emotionally perceived each visualization, and in so doing, we investigate our \emph{RQ}. To answer this question, we turned to crowdsourcing.

\begin{figure}
\centering
  \includegraphics[width=0.9\linewidth]{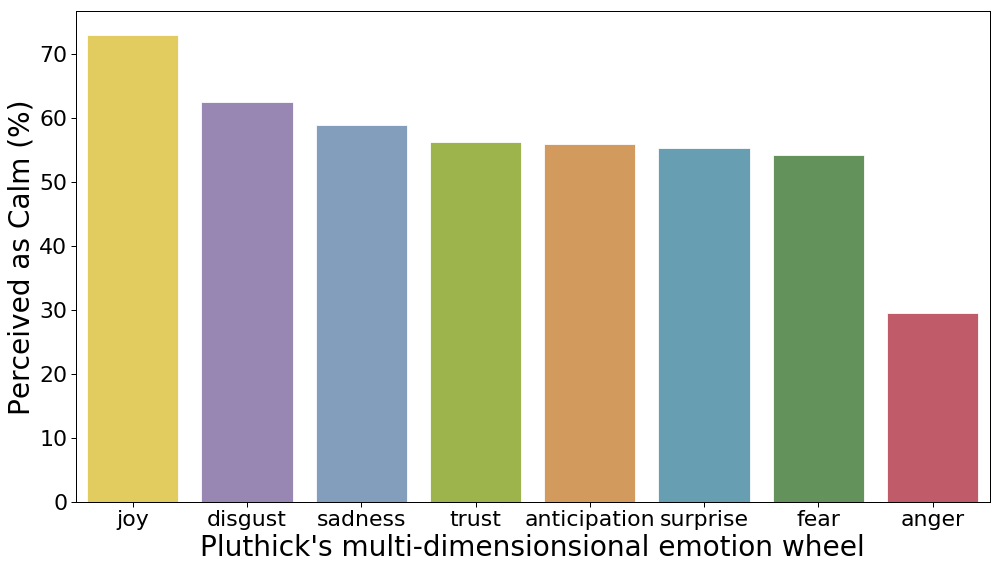}
  \caption{Calmness levels of the eight collective emotional states visualizations as our participants perceived them (responses to $Q_{2}$). Consistently, as one would expect, joy was perceived as the most relaxing one, while anger and fear were perceived as less relaxing.}
 
\label{fig:calm_levels}
\end{figure}

\subsection{Online Study}

\subsubsection{Participants and procedure}
We made HeartBees available on Amazon Mechanical Turk (AMT). We ensured to enrol highly reputable AMT workers by targeting workers with 95\% HIT approval rate and at least 100 approved HITs, and we received a total of 353 unique responses (each worker evaluated HeartBees once). The online workers were presented with a series of dynamic boids visualizations representing the eight emotions (Table~\ref{tbl:boids_coefficients}). Their task was to evaluate which emotions each visualization conveyed and assess its calmness levels. In particular, each worker responded to two questions for each visualization. The first question probed the extent to which each emotional state was easily recognizable in the boids configuration, while the second question explored whether our participants perceived the different visualizations of collective emotional states to be calm, and served as an orthogonal check to the first question. The two questions were as follows:

\begin{description}
    \item $Q_{1}$: Which emotion does the visualization trigger? \\Possible Answer: Radio button selector with the eight emotions
    \item $Q_{2}$: Does the visualization make you feel relaxed? \\ Possible Answer: Likert scale 1-5 (1: Strongly Disagree, 5: Strongly Agree)
\end{description}

To mitigate possible interface biases, we reshuffled the order with which the eight visualizations of the collective emotional states are presented as well as the potential responses. To make sure we kept only responses in which workers spent a fairly sufficient time to make a judgment, we removed those responses that took less than 7 seconds on average~\cite{first_impression}. The resulting dataset consisted of 340 responses (49\% female, $\mu_{age} = 33.1$, $\sigma_{age}=9.8$). 

\begin{table*}
\centering
  \caption{Matrix of emotion co-occurrences. \emph{Top:} Normalized responses to $Q_{1}$ (columns represent the users' associations, and rows represent the eight emotions from our visualizations). 
  \emph{Bottom:} Normalized emotions in Flickr data (columns represent the users' associations, and rows represent emotions extracted using the Emolex dictionary).
  }
  \label{tab:results_q1}
  \begin{tabular}{cccccccccc}
  \toprule
    & & \multicolumn{8}{c}{\textbf{User Associations}} \\
    \toprule
      & & \textbf{Joy} & \textbf{Disgust} & \textbf{Trust} & \textbf{Anger} & \textbf{Surprise} & \textbf{Anticipation} & \textbf{Sadness} & \textbf{Fear}\\
    \midrule
     \multirow{8}{*}{\rotatebox[origin=c]{90}{Emotions from viz}} & \textbf{Joy} & \textbf{0.17} & 0.03 & 0.15 & 0.05 & 0.09 & 0.14 & 0.11 & 0.06 \\
     & \textbf{Disgust} & 0.14 & 0.07 & 0.11 & 0.06 & 0.12 & 0.14 & 0.12 & 0.12\\
     & \textbf{Trust} & 0.11 & 0.07 & 0.17 & 0.14 & 0.09 & \textbf{0.15} & 0.10 & 0.13 \\
     & \textbf{Anger} & 0.07 & 0.10 & 0.03 & \textbf{0.40} & \textbf{0.27} & 0.09 & 0.03 & \textbf{0.22}\\
     & \textbf{Surprise} & 0.15 & 0.08 & 0.12 & 0.06 & 0.09 & \textbf{0.15} & 0.10 & 0.09 \\
     &\textbf{Anticipation} & 0.16 & 0.10 & 0.09 & 0.08 & 0.15 & 0.11 & 0.08 & 0.10\\
     & \textbf{Sadness} & 0.07 & \textbf{0.34} & \textbf{0.21} & 0.06 & 0.08 & 0.09 & \textbf{0.28} & 0.13\\
     & \textbf{Fear} & 0.12 & 0.19 & 0.12 & 0.13 & 0.10 & 0.12 & 0.15 & 0.13 \\
     \midrule
     \midrule
     \multirow{8}{*}{\rotatebox[origin=c]{90}{Emotions from Flickr}} & \textbf{Joy} & 0.09 & 0.02 & \textbf{0.12} & 0.03 & \textbf{0.07} & \textbf{0.14} & 0.06 & 0.04\\
     & \textbf{Disgust} & 0.12 & 0.02 & 0.09 & 0.06 & 0.05 & 0.10 & 0.08 & 0.07\\
     & \textbf{Trust} & \textbf{0.16} & 0.02 & 0.06 & 0.04 & 0.06 & 0.13 & 0.07 & 0.05\\
     & \textbf{Anger} & 0.11 & \textbf{0.03} & 0.10 & 0.03 & 0.05 & 0.10 & 0.08 & \textbf{0.09} \\
     & \textbf{Surprise} & \textbf{0.16} & 0.02 & 0.11 & 0.03 & 0.03 & 0.13 & 0.07 & 0.05\\
     & \textbf{Anticipation} & \textbf{0.16} & 0.02 & \textbf{0.12} & 0.03 & \textbf{0.07} & 0.06 & 0.07 & 0.04 \\
     & \textbf{Sadness} & 0.13 & \textbf{0.03} & 0.11 & 0.05 & 0.06 & 0.11 & 0.04 & 0.07 \\
     & \textbf{Fear} & 0.10 & \textbf{0.03} & 0.10 & \textbf{0.07} & 0.05 & 0.10 & \textbf{0.09} & 0.05 \\
    \bottomrule
  \end{tabular}
\end{table*}

\subsection{Analysis}
To investigate the extent to which the emotion configurations encoded in the Boids model matched people's perceptions, we analyzed  responses to $Q_{1}$. To ensure comparable results across the eight emotions, we normalized our participants' responses by dividing the number of occurrences for each response (i.e., one of the eight available options of $Q_{1}$) with the total number of occurrences for that response. For the sake of comparing the resulting associations between displayed and perceived emotions, we also gathered data about typical associations between basic emotions found online. \revision{This allowed us to assess whether the confusion that people had in identifying some of the emotions could be due to the high similarity between those emotions in the way people conceptualize them.} Specifically, we used the 100M photos from the Flickr Creative Commons public dataset~\cite{thomee2016yfcc100m}. From the set of all the photos in the dataset, we collected the user-generated tags attached to each picture and we matched them against the EmoLex dictionary~\cite{mohammad2013crowdsourcing} to extract only those that express any of the eight basic emotions. We then counted the number of times tags expressing emotion $i$ co-occur in the same picture with tags expressing emotion $j$, for all pairs of emotions. We apply the same normalization we used for the crowdsourcing responses to these co-occurrence counts.

\subsection{Results}
Results of the crowdsourcing task to $Q_{1}$ and the external validity check using Flickr data are summarized in Table~\ref{tab:results_q1}. The Boids models expressing Joy, Anger, and Sadness were correctly identified by the crowdworkers. Most often, the workers associated fear and disgust respectively to the visualizations that were supposed to convey anger and sadness. These specific pairs of emotions are associated frequently also in the Flickr data, which signals the fact that they might be frequently associated in the human mind; interesting emotional states that fall into the negative spectrum of the emotion wheel. Trust and surprise were not matched correctly, as they were associated most often to the sadness and anger visualizations, respectively. To a certain extent, the same picture was observed in Flickr data as no distinct patterns emerged. Users often associated them not only with positive ones (e.g. associate trust and surprise with joy), but also signals that suggest association with sadness and anger respectively were observed.

Results of the crowdsourcing task related to $Q_{2}$ are shown in Figure~\ref{fig:calm_levels}. We found that, as one would expect, the boids configuration which conveys joy was the one perceived as the most relaxing one with 72\% agreement among our participants' responses. On the contrary, the anger configuration was perceived as the least relaxing one with only 29\% agreement, followed by the configuration expressing fear. Trust, surprise, and anticipation were neutral, but skewed towards calmness, while sadness and disgust shared a slightly even higher skewness towards agreement with relaxation levels.

\section{Discussion and Conclusion}
\label{sec:section7}

Emotional Intelligence is a valuable asset for both individuals and groups. Mechanisms to foster it promote well-being and productivity. We proposed HeartBees, a bio-feedback system for visualizing collective emotional states through the use of a nature-inspired metaphor. Drawing inspiration from observational studies on animals behaviour, HeartBees simulates a flock of birds that create coordinated patterns. These patterns create a visual metaphor of the complexities of our emotional states, and they do so in a collective way.  

We evaluated our system's ability to convey the multi-faceted nature of emotions in a crowdsourcing task with 353 participants. We found that consensus emerged among participants' subjective perceptions for three emotions. Negative emotions were broadly recognized, perhaps because of human's inherent ability of perceiving negative emotions better than positive ones~\cite{vaish2008not}. The ability to effectively convey the presence of negative emotions through visuals provides a useful mean to raise awareness about the distress of groups, and possibly trigger appropriate rebalancing interventions.

Our work has both theoretical and practical implications. From the theoretical standpoint, it offers a framework to integrate concepts and metaphors in a new way of visualizing complex data. We showed, for example, how complex and unconventional data streams such as biosignals could be transformed into abstract, relatable, and universally accessible visuals to convey our collective emotional states. Additionally, our system contributes to the emerging field of Inbodied Interactions~\cite{schraefel2019in5}, which foresees future technologies to better align with how our body internally works. From a practical perspective, we foresee that our system could be deployed in a number of settings, including the workplace. This would provide immediate feedback of the workplace's atmosphere and allow managers to assess the `health' of their teams (e.g., stress levels). As the overarching goal of the visualization was to incorporate elements of collective behavior, HeartBees would allow co-workers to experience synchronicity between them as expressed through their emotions. In so doing, it would nudge people into experiencing more positive emotions. A practical way of deploying such a system in the workplace would be an interactive installation in which the Boids behavior is controlled by the co-workers or by-passers who wear wearable devices such as smartwatches. Importantly, the collective way of visualizing people's emotional states could be deemed appropriate in the workplace due to the privacy-preserving nature of HeartBees. This means that no individual emotions could be identified as our system visualizes data in a collective way.

This work has \revision{two limitations that suggest directions for future work}. While high degrees of consensus emerged among the ways our participants subjectively perceived the visualizations, \revision{future studies could further evaluate the end-to-end system in a participatory environment in real time.} Future studies could also further fine-tune the motion configurations of the Boids model.  However, our adjustable interface provides the means of direct manipulation of those parameters, thus allowing practitioners to make use of our solution and experiment with it. \revision{The second limitation  concerns the aesthetics of the Boids visualization. In future studies, we plan to investigate the role of color, texture, and quantity of objects on screen in conveying the eight emotional states. Furthermore, the way people perceive color depend on  gender, cultural background, or personal experiences; thus future studies could investigate the role of these factors too.} Future directions also include the exploration of alternative modalities for conveying emotions, including the sonification of HeartBees, not least because the use of particular animals' sounds have been linked to increased laughter in people~\cite{schwing2017positive}.

% The authors wish to thank A, B, and C. This work was supported in part by
% a grant from XYZ (\# 12345-67890).}

%\bibliographystyle{abbrv}
\bibliographystyle{abbrv-doi}

\bibliography{references}
\end{document}